\documentclass[a4paper]{panl}
\usepackage{cite}
\usepackage{wrapfig}
\usepackage{graphicx}
\usepackage{amssymb}
\usepackage{amsfonts}
\usepackage{amsmath}
\usepackage{longtable}
\usepackage{rotating}
\usepackage{lscape}
\usepackage{epsfig}
\usepackage{multirow}
\usepackage{xspace}
\usepackage{lineno}

\newcommand*{\kT}{\ensuremath{k_\mathrm{T}}\xspace}
\newcommand*{\pTjet}{\ensuremath{p_\mathrm{T}^{\mathrm{jet}}}\xspace}
\newcommand{\pTchjet}{\ensuremath{p_\mathrm{T,ch\,jet}}\xspace}

\newcommand*{\Lcp}{\ensuremath{\Lambda_\mathrm{c}^+}\xspace}
\newcommand*{\theg}{\ensuremath{\theta_\mathrm{g}}\xspace}
\newcommand*{\zg}{\ensuremath{z_\mathrm{g}}\xspace}

\originalTeX
\begin{document}

\title{Jet substructure measurements with ALICE}
\maketitle
\authors{R.\,Vértesi$^{a,}$\footnote{E-mail: vertesi.robert@wigner.hu} (for the ALICE Collaboration)}
\setcounter{footnote}{0}
\from{$^{a}$\,Wigner Research Centre for Physics, MTA Centre of Excellence, \\
	29-33 Konkoly-Thege Miklós út, 1121 Budapest, Hungary}

\begin{abstract}
A selection of new jet substructure measurements are reported from the ALICE experiment at the CERN LHC in both proton-proton and heavy-ion collisions. These include the first fully corrected inclusive measurements of the groomed jet momentum fraction and the groomed jet radius, as well as the $N$-subjettiness distribution and the fragmentation distribution of reclustered subjets. We also report on the measurement of several groomed substructure observables of heavy-flavor jets in pp collisions, fragmentation functions and the new measurements of the radial distributions of D$^0$ mesons or $\Lambda^+_c$ baryons in jets. The measurements are compared to theoretical calculations and provide new constraints on the physics underlying parton fragmentation and jet quenching.
\end{abstract}
\vspace*{6pt}


\label{sec:intro}
\section{Introduction}

Jet substructure measurements, based on the distribution of constituents within a jet, are able to probe specific regions of QCD radiation phase space for jet showers in vacuum. This powerful capability provides new opportunities to study fragmentation patterns of parton showers in vacuum and the dynamics of jet quenching in heavy-ion collisions.

The ALICE detector~\cite{ALICE:2014sbx} has unique capabilities for jet substructure measurements, due to its high-precision tracking system and focus on jets with low transverse momenta. 
Charged-particle jets are clusterized using the anti-\kT algorithm~\cite{Cacciari:2008gp} from tracks that are reconstructed in the Inner Tracking System and the Time Projection Chamber, with a high spatial precision in the full azimuth angle $\varphi$ and the pseudorapidity range $|\eta|<0.9$.
Full jets reconstructed using the Electromagnetic Calorimeter in a more limited acceptance ($1.4<\varphi<\pi$, $|\eta|<0.7$) allow for a more direct access to parton kinematics.

Heavy-flavor hadrons are identified from their decay products, utilizing tagging algorithms based on the displacement of the secondary decay vertex.
The excellent tracking capabilities of the ALICE detector allow for jet substructure studies for heavy-flavor as well as untagged jets.
Heavy-flavour jets are declustered to trace all branchings of the charm quark and to reveal mass dependence of the shape and structure of the parton shower due to the dead-cone effect~\cite{Cunqueiro:2018jbh}.

In the following sections, a selection of recent ALICE results is presented on inclusive untagged-jet and charm-jet substructure observables in proton-proton collisions, followed by results on jet substucture in heavy-ion collisions. 

\label{sec:results}

\section{Groomed jet substructure of inclusive jets in pp collisions}

Jet grooming algorithms are used to access the hard parton structure of a jet. By mitigating the influence from the underlying event and hadronization processes, perturbative quantum chromodynamics (pQCD) calculations can be directly compared to groomed-jet observables. A popular novel technique, soft-drop (SD) grooming~\cite{Larkoski:2014wba}, aims for the removal of large-angle soft radiation. 
In this method, the jets that had previously been reconstructed with the anti-\kT algorithm~\cite{Cacciari:2008gp} are reclustered using the Cambridge-Aachen (C/A) algorithm~\cite{Dokshitzer:1997in} to form a clustering tree that follows angular ordering. Then the soft branches are iteratively removed if not fulfilling the so-called soft-drop condition, 
\begin{eqnarray}
z > z_{\rm cut} \theta^\beta\,, & {\rm where} &  z = \frac{p_{{\rm T},2}}{p_{{\rm T},1}+p_{{\rm T},2}} \ {\rm and} \ \theta = \frac{\Delta R_{1,2} }{ R } 
\end{eqnarray}
are the momentum fraction taken by the subleading prong ($p_{{\rm T},1}$ and $p_{{\rm T},2}$ being the momenta of the two prongs), and the splitting radius (defined as the ratio of the $\Delta R_{1,2}$ splitting angle between the two prongs and the resolution parameter $R$ of the anti-\kT clustering). The soft threshold $z_{\rm cut}$ determines the overall strength of the grooming, and the angular exponent $\beta$ 
controls the strength with which wide-angle soft radiation is rejected.
The groomed-jet substructure is often characterized by the groomed momentum fraction $\zg$ and the groomed radius $\theg$, defined as the values of $z$ and $\theta$ corresponding to the first hard splitting fulfilling the soft-drop condition. In addition, the number of groomed-jet splittings fulfilling the SD-condition along the hardest branch, $n_{\rm SD}$, is also used to quantify the jet substructures.
ALICE measurements of charged-particle-jet $\theg$, $\zg$ and $n_{\rm SD}$ in pp collisions are well-described by model calculations~\cite{ALICE:2019ykw,Vertesi:2020gqr}.

Jet structure can also be characterized with the generalized jet angularities
\begin{eqnarray}
\lambda_\alpha^\kappa = \sum_{i} z_i^\kappa \theta_i^\alpha\,, & {\rm where} & z_i = \frac{p_{{\rm T},i}}{\pTjet} \ {\rm and} \ \theta_i = \frac{\Delta R_i}{R}
\end{eqnarray}
are the momentum fraction and angular deflection of the $i^{\rm th}$ constituent within the jet~\cite{Larkoski:2014pca}.
The generalized jet angularities are infrared- and collinear-safe quantities in case of $\kappa=1$ and $\alpha>0$, in which case they can be directly calculated from pQCD.
The special cases $\lambda_1^1$ and $\lambda_2^1$ reduce to the jet girth and jet thrust. 
With the comparison of $\lambda_\alpha^\kappa$ for groomed and non-groomed jets, with the systematic variation of $\alpha$, the interplay between the perturbative and non-perturbative regimes of the QCD can be explored in detail, and strong constraints can be established for pQCD and fragmentation models.

\begin{figure}[h]
	\begin{center}
		\vspace{-4mm}
		\includegraphics[width=.9\textwidth]{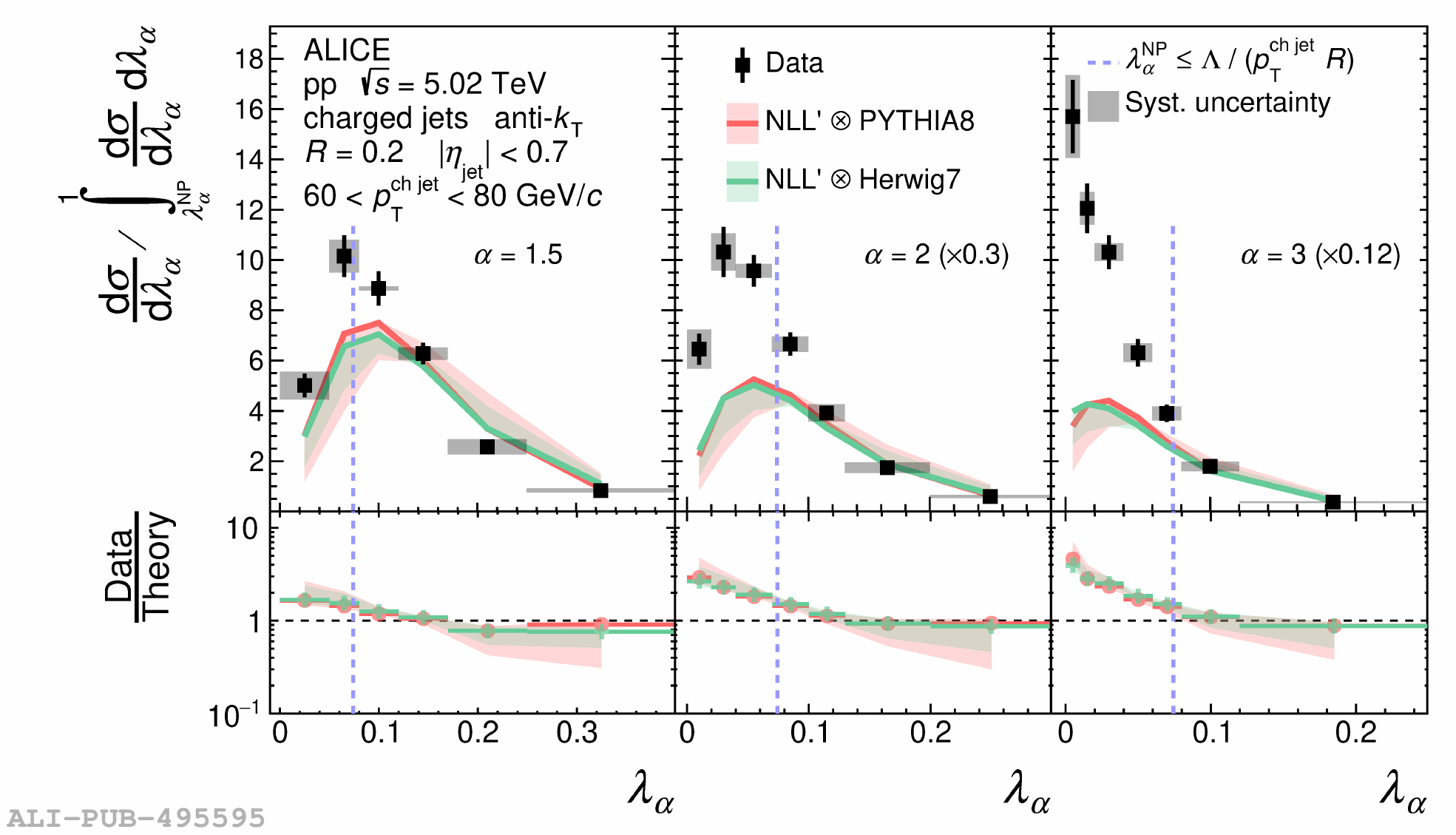}		\includegraphics[width=.9\textwidth]{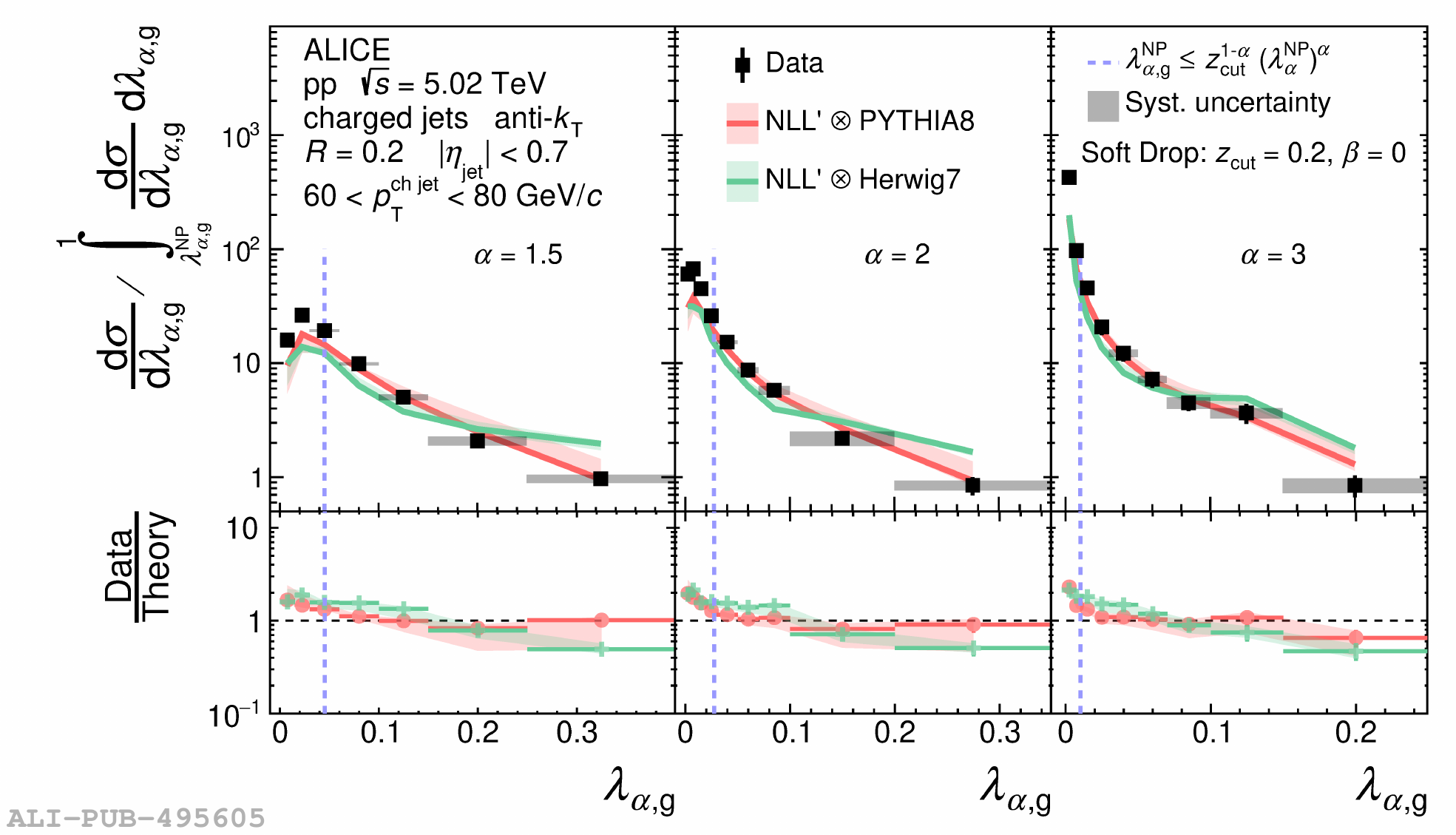}
		\vspace{-4mm}
		\caption{Comparison of ungroomed (top) and groomed (bottom) charged-particle jet angularities in pp collisions for $R = 0.2$ to analytical NLL$^\prime$ predictions with MC hadronization corrections in the range $60<\pTchjet<80$ GeV/$c$~\cite{ALICE:2021njq}. The dashed blue line indicates the boundary between the non-perturbative and perturbative regimes.}
	\end{center}
	\labelf{fig:Angularity}
	\vspace{-5mm}
\end{figure}

Figure \ref{fig:Angularity} shows the generalized jet angularities measured by the ALICE experiment in pp collisions at $\sqrt{s}=5.02$ TeV for different $\alpha$ values for ungroomed (top) and SD-groomed jets (bottom)~\cite{ALICE:2021njq}. The data are also compared to next-to-leading-logarithmic (NLL$^\prime$) calculations~\cite{Almeida:2014uva} with PYTHIA8~\cite{Sjostrand:2007gs} as well as Herwig7~\cite{Bellm:2015jjp} fragmentation. In the ungroomed case, a significant deviation can be observed in the non-perturbative regime, toward smaller $\lambda_\alpha$ values. The deviation is more pronounced in the case the $\alpha$ exponent is larger. As expected, the data is better reproduced by the models in the groomed case, as the perturbative regime is significantly extended by SD grooming.

\section{Flavor dependence of the jet structure}

Due to their large mass, heavy-flavor hadrons are produced in perturbative processes down to low transverse momenta. Since heavy-flavor quarks undergo weak decay, their numbers remain largely unchanged throughout the later evolution of the jets. Identification of jets initiated by heavy-flavor quarks therefore allow for direct access to jets initiated by quark fragmentation. Jet fragmentation is flavor dependent 
due to the color charge of the initial parton and also due to quark-mass effects on the radiation pattern. In the LHC energy regime, most jets originate from gluon fragmentation and are typically softer and wider than quark jets due to their larger color charge~\cite{Kluth:2006bw}. Charged particles with a mass $m>0$ and energy $E$ emit radiation that is suppressed below angles $\theta \approx m/E$ with respect to the axis of the radiator. This so-called dead-cone effect is expected to be present in jets containing heavy flavor~\cite{Dokshitzer:1991fd,Thomas:2004ie}. As a consequence, heavy-flavor quarks fragment hard, meaning that on average, radiation by the heavy parton carries a smaller fraction of the momentum than in the case of light partons.
The ALICE Collaboration recently reported the first direct observation of the QCD dead-cone by using iterative declustering techniques to reconstruct the parton shower of charm quarks~\cite{ALICE:2021aqk}.

\begin{figure}
	\centering
	\includegraphics[width=0.33\textwidth]{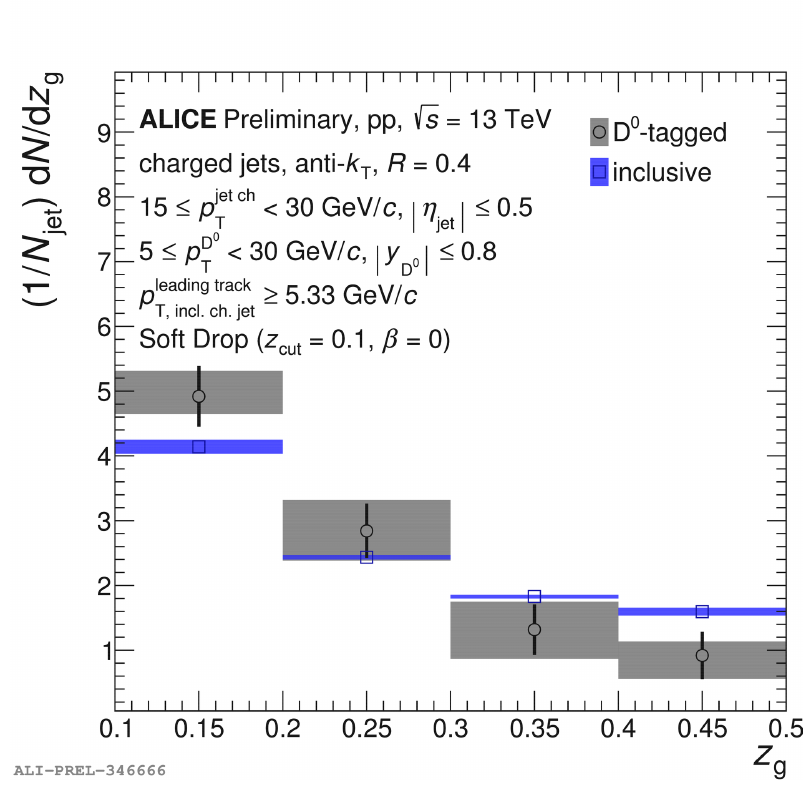}%
	\includegraphics[width=0.33\textwidth]{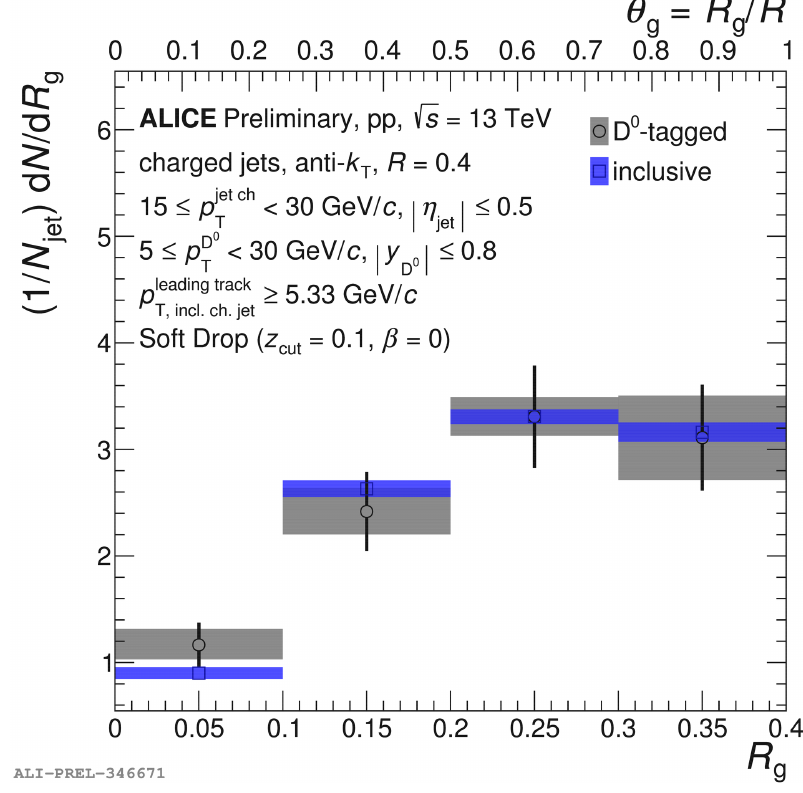}%
	\includegraphics[width=0.33\textwidth]{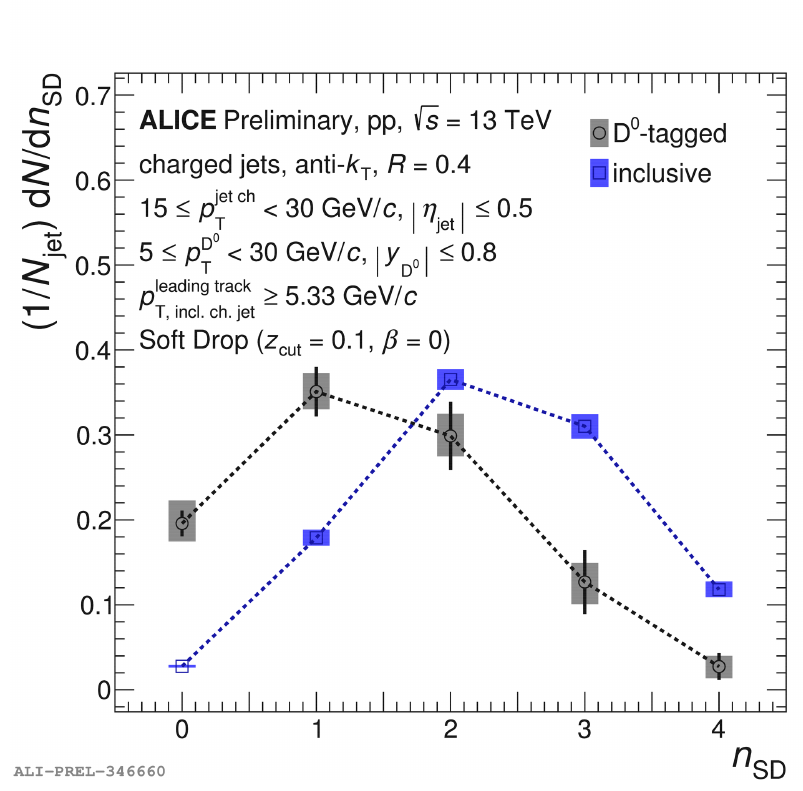}%
	\caption{\label{fig:dtagsubstr}Substructure variables $\zg$ (left), $\theg$ (center) and $n_{\rm SD}$ (right) of D$^0$-tagged charged-particle jets compared to inclusive charged-particle jets, in pp collisions at $\sqrt{s}=13$ TeV.}
\end{figure}

The consequences of the dead-cone effect on heavy flavor jet fragmentation can be also seen in 
Fig.~\ref{fig:dtagsubstr}, which shows the first measurement of groomed jet substructure of D$^0$-tagged jets, compared to that of inclusive jets, in pp collisions at $\sqrt{s}=13$ TeV.
While the slightly different trends in $\zg$ and $\theg$ between charmed and inclusive jets give a hint about flavor-dependent jet substructure, on average, there is about one less hard soft-drop splitting in charm jets than in inclusive jets. This fact is consistent with harder heavy-flavor fragmentation caused by the massive quarks.

\begin{figure}
	\centering
	\includegraphics[width=0.5\textwidth]{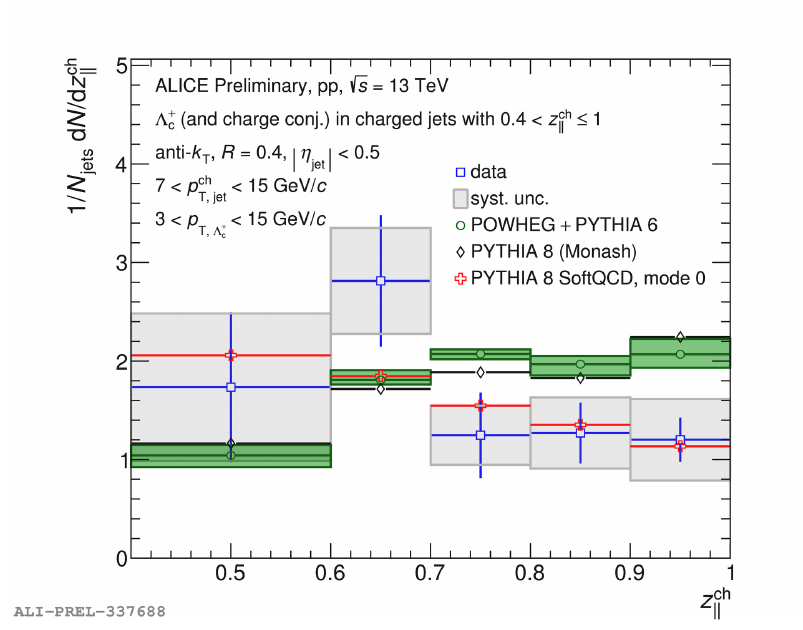}%
	\includegraphics[width=0.5\textwidth]{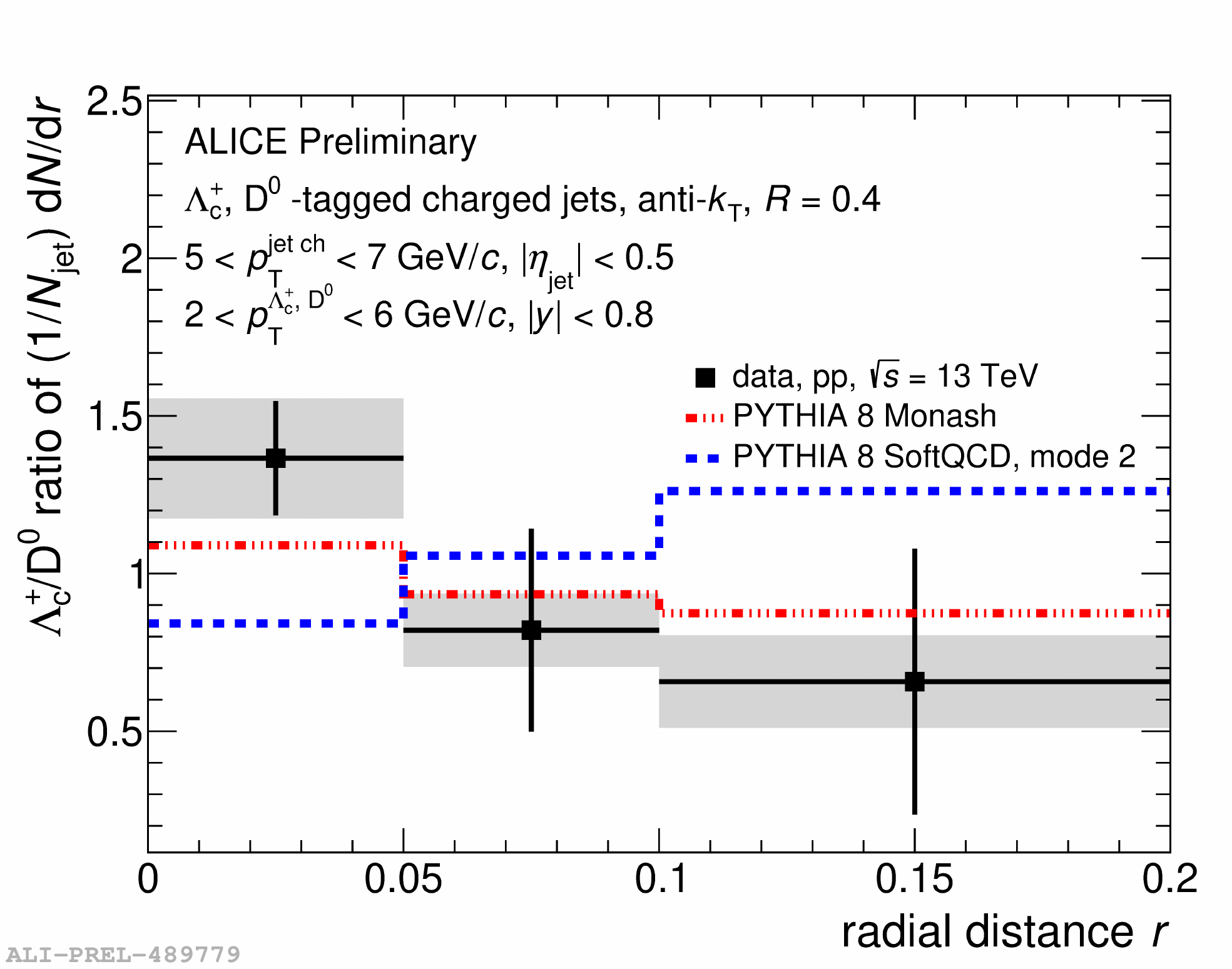}%
	\caption{Parallel momentum fraction $z_{\parallel}$ of charged-particle jets tagged with \Lcp baryons (left), and the $\Lambda_c^+/{\rm D}^0$ ratio in function of angular distances (right). Both are compared to PYTHIA8 with the Monash tune~\cite{Sjostrand:2007gs} as well as the enhanced color-reconnection scenario~\cite{Christiansen:2015yqa}.}
	\label{fig:zpar}
\end{figure}

While the production of heavy-flavor mesons via fragmentation follows a common pattern in different systems, recent measurements show that this universality is broken in the baryonic sector~\cite{ALICE:2020wfu,ALICE:2021bli}. Although several theoretical scenarios are proposed, including enhanced color reconnection with color-string junctions, statistical hadronization with enhanced baryonic states as well as quark coalescence~\cite{Christiansen:2015yqa,He:2019tik,Plumari:2017ntm}, the proper description of charmed-baryon production still remains a challenge.
The reconstruction of heavy-flavor hadrons within a jet allows for direct access to the fragmentation of charm or beauty quark without the need to reconstruct the groomed jet substructure. ALICE measures both the parallel momentum fraction $z_{\parallel}$ carried by the heavy-flavor hadron from the jet, and the radial angular distance of the heavy-flavor hadron from the jet axis for both the charmed D$^0$ mesons and the $\Lambda_c^+$ baryons. Figure~\ref{fig:zpar} shows $z_{\parallel}$ distribution for jets with $\Lambda_c^+$, as well as the $\Lambda_c^+/{\rm D}^0$ ratio, compared to models.

While the $z_{\parallel}$ shows similar longitudinal behavior to the case of ${\rm D}^0$ mesons~\cite{Vertesi:2020gqr}, the radial-distance dependent ratios suggest that on the average, the $\Lambda_c^+$ fragments closer to the jet axis than the ${\rm D}^0$. Interestingly, neither of the trends are reproduced by models assuming enhanced color reconnection with color-string junctions~\cite{Christiansen:2015yqa}.

\section{Jet substructure in heavy-ion collisions}

In collisions of ultrarelativistic heavy-ions, jet substructure measurements access the modification of jet fragmentation by the deconfined medium. 
Figure~\ref{fig:HeavyIonSD} shows the fully unfolded groomed momentum fraction $\zg$ and the groomed radius $\theg$ of $R=0.2$ jets in Pb--Pb compared to pp collisions at $\sqrt{s_{\rm NN}}=5.02$ TeV, as well as their ratios compared to several model calculations. 
The combinatorial background is suppressed using event-wise constituent subtraction, and a strong grooming condition of $z_{\rm cut}=0.2$ is applied.
\begin{figure}[h]
	\begin{center}
		\includegraphics[width=0.5\textwidth]{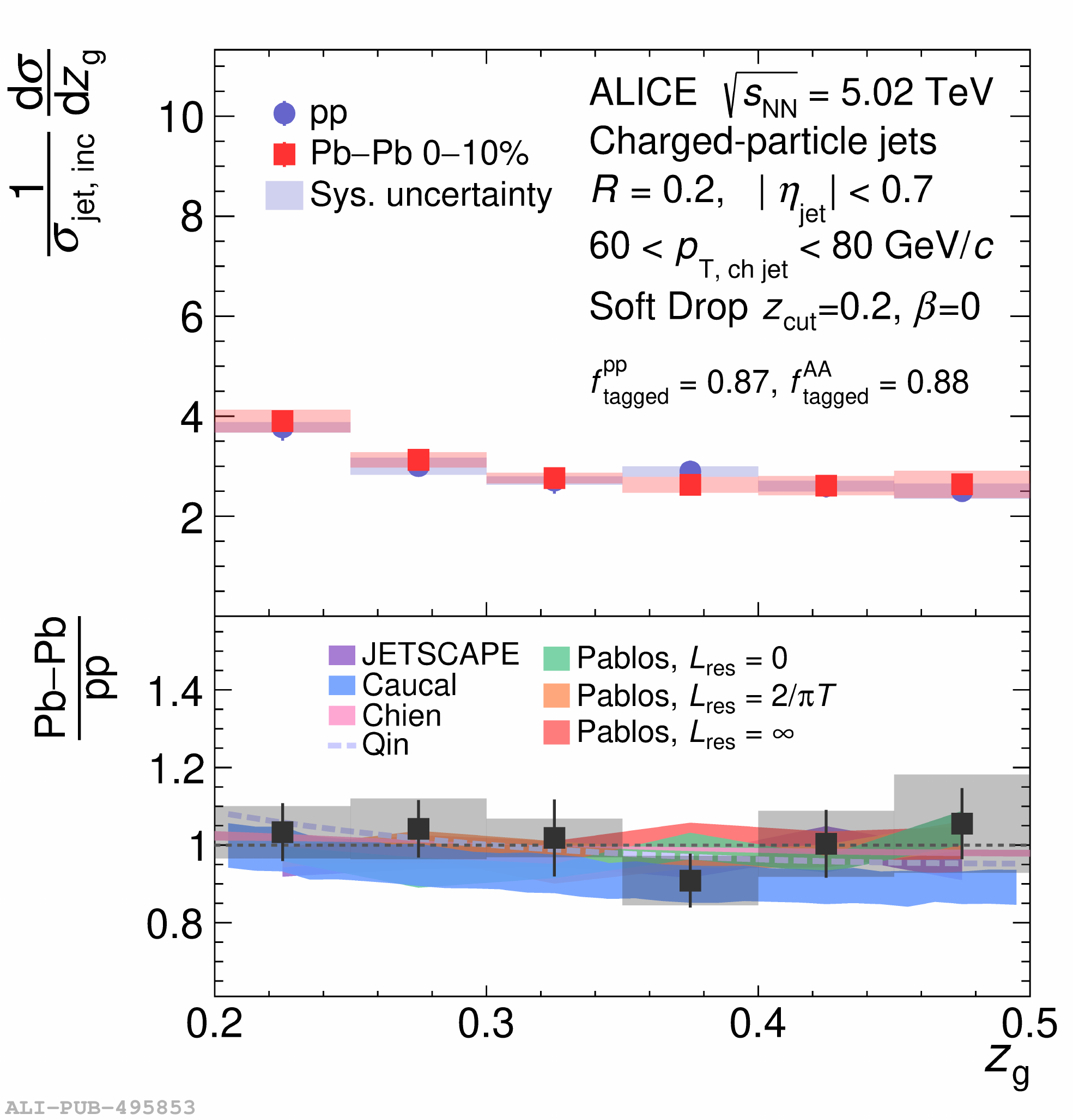}%
		\includegraphics[width=0.5\textwidth]{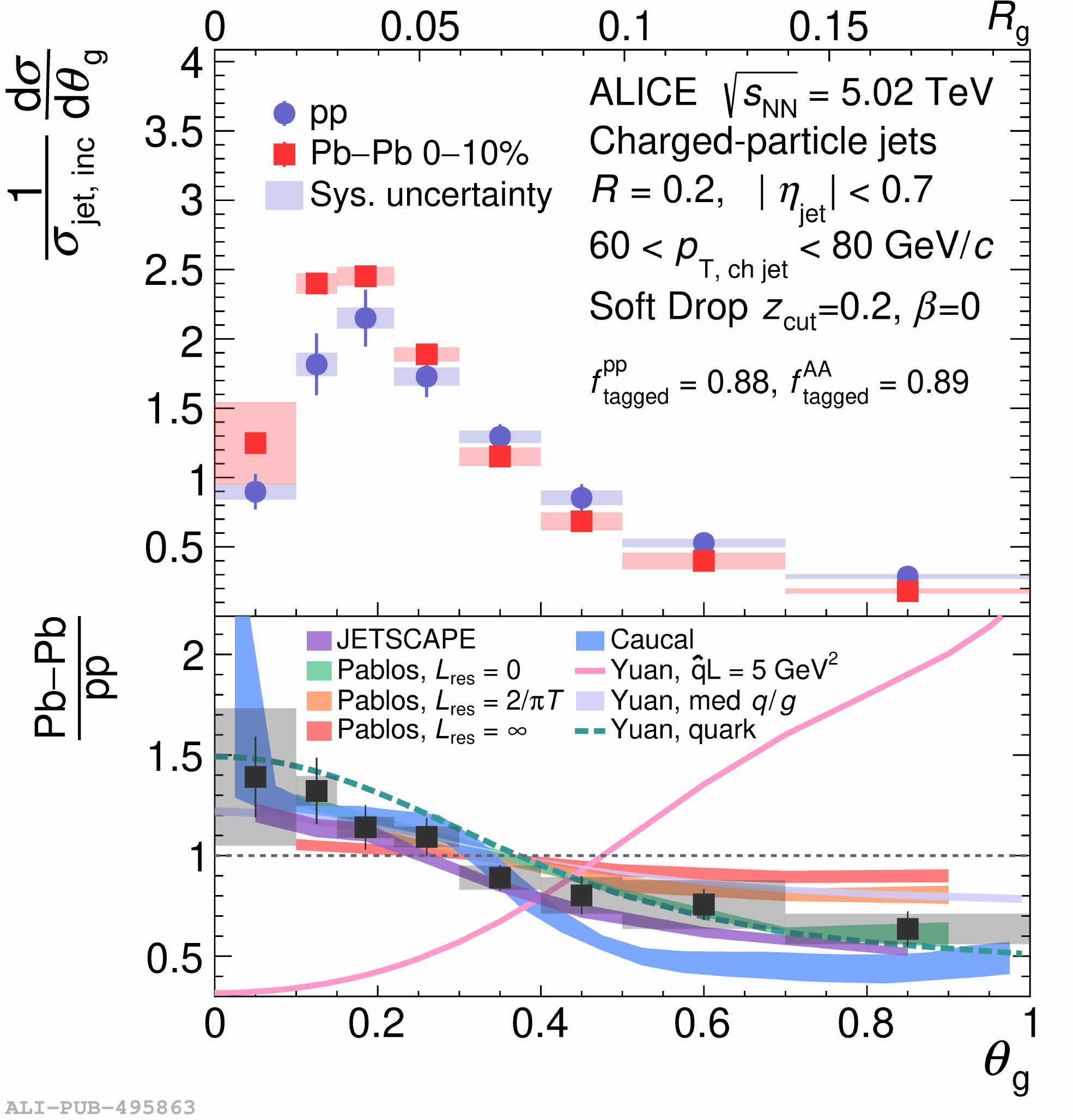}
		\vspace{-3mm}
		\caption{Unfolded $\zg$ (left) and $\theg$ distributions (right) for charged-particle jets in pp collisions compared to those in Pb–Pb collisions at $\sqrt{s_{\rm NN}}=5.02$ TeV with $z_{\rm cut} = 0.2$ for 0–10\% centrality and $R = 0.2$, in the $60<\pTchjet<80$ GeV/$c$ jet transverse-momentum range~\cite{ALICE:2021obz}.
	}
	\end{center}
	\labelf{fig:HeavyIonSD}
	\vspace{-5mm}
\end{figure}
The distributions in $\zg$, representing the structure parallel to the jet axis, show no modification within uncertainties, The transverse quantity $\theg$, however, displays a suppression of large angles, and an enhancement of small angles, hinting that the medium filters out wider subjets. Models with incoherent energy loss as well as fully coherent energy loss with gluon filtering qualitatively describe the data (see Ref.~\cite{ALICE:2021obz} and references therein).

Another way to access the details of radiation patterns is to construct observables that probe subjets (or more closely collimated prongs of particles) within a jet.
One such class of variables is $N$-subjettiness~\cite{Thaler:2010tr}, defined as
\begin{eqnarray}
\tau_N = \frac{1}{\pTjet R}\sum_k p_{{\rm T},k}{\rm min}(\Delta R_{1,k}, \Delta R_{2,k}, ..\Delta R_{N,k})\,,
\end{eqnarray}
where $k$ runs over the constituents of the jet, and $\Delta R_{i,k}$ are the distances of each subjet candidate $1 \le i \le N$ and the constituent $k$.
By construction $\tau_N \approx 1$ in case the number of subjet prongs is less than $N$, and $\tau_N \approx 0$ otherwise. The distribution of $\tau_2 / \tau_1$ will therefore be sensitive to the occurrence of 2-pronged vs.\ 1-pronged jets. 
Subjet fragmentation can also be accessed by reclustering jets using a smaller resolution parameter, $r<R$, and then characterize leading subjets with momentum fraction
\begin{eqnarray}
z_r = \frac{p_{\rm}^{\rm leading\ subjet}}{p_{\rm}^{\rm jet}}\,.
\end{eqnarray}

\begin{figure}[h]
	\begin{center}
		\includegraphics[width=0.52\textwidth]{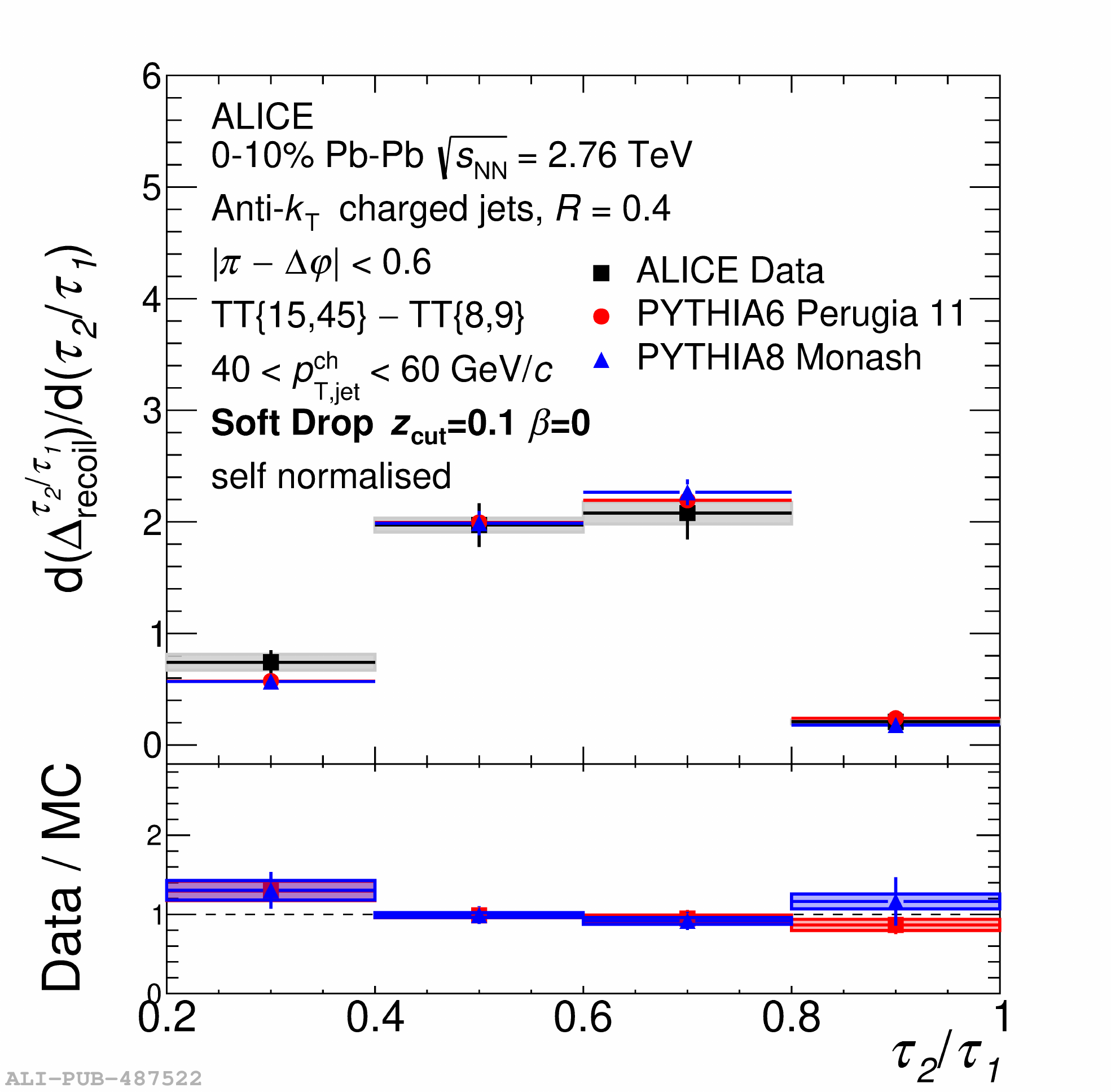}%
		\includegraphics[width=0.48\textwidth]{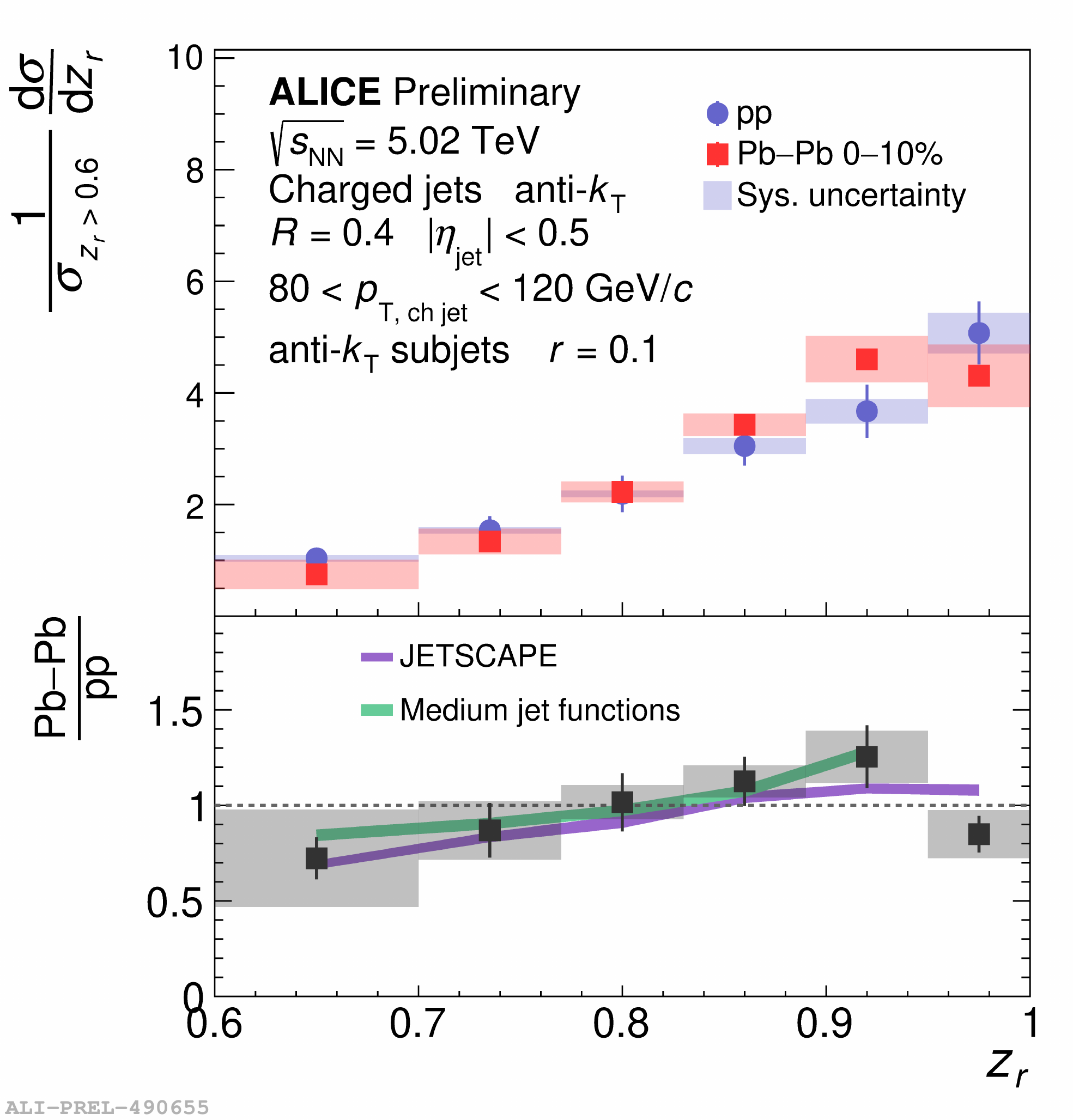}
		\vspace{-3mm}
		\caption{Left: fully corrected $\tau_2/\tau_1$ distributions with SD grooming in Pb--Pb collisions at $\sqrt{s_{\rm NN}}=2.76$ TeV for charged-hadron jets with $R=0.4$ in the $40<\pTchjet<60$ GeV/$c$ range, compared to models~\cite{ALICE:2021vrw}. Right: Subjet fragmentation $z_r$ in Pb--Pb collisions at $\sqrt{s_{\rm NN}}=5.02$ TeV in  $80<\pTchjet<120$ GeV/$c$ range, compared to models.}
	\end{center}
	\labelf{fig:HeavyIonSubjets}
	\vspace{-5mm}
\end{figure}
Figure~\ref{fig:HeavyIonSubjets} (left) shows the first measurement of $\tau_2/\tau_1$ distributions, in Pb--Pb collisions at $\sqrt{s_{\rm NN}}=2.76$ TeV, with C/A reclustering and SD grooming~\cite{ALICE:2021vrw}. The data show no significant modification within the current precision compared to PYTHIA6 and 8 simulations~\cite{Sjostrand:2006za,Sjostrand:2007gs}.
In Figure~\ref{fig:HeavyIonSubjets} (right), the subjet fragmentation $z_r$ is shown in Pb–Pb collisions at $\sqrt{s_{\rm NN}} = 5.02$ TeV and a subjet resolution parameter $r=0.1$. Consistently with model predictions, the data show hints of medium modification in the quark-dominated range around $z_r\approx 1$.

\section{Summary}

This contribution summarized some of the recent results on jet substructure observables from the ALICE Collaboration. Measurements in pp collisions primarily serve to test predictions of pQCD calculations and hadronization models. Hard pQCD processes can be separated from soft radiation using grooming techniques.
Flavor-dependent jet substructure measurements can be used to explore heavy-flavor fragmentation and help in disentangling color-charge and mass effects.
Groomed jet substructures provide possibility to explore flavor and mass-dependent fragmentation. ALICE presented the first direct observation of the dead cone in hadronic collisions and demonstrated harder heavy-flavor fragmentation in groomed ${\rm D}^0$-tagged jets.
Jet substructure measurements in Pb--Pb collisions aim for the understanding of jet modification by the hot and dense deconfined medium. 

The upcoming LHC Run--3 data-taking phase with higher luminosity~\cite{Noferini:2018are} will allow for precision measurements of beauty-jet substructure as well as detailed measurements in the charm baryonic sector. These will allow for the separation of color-charge and mass effects as well as the understanding of fragmentation in details. Future heavy-ion measurements can explore the details of jet--medium interaction, further facilitating model development and moving toward a deeper understanding of the non-perturbative domain of the strong interaction.


This work has been supported by the Hungarian NKFIH/OTKA FK 131979 and K 135515 grants, as well as the NKFIH 2019-2.1.6-NEMZ\_KI-2019-00011 project.

\bibliographystyle{pepan}
\bibliography{biblio}

\end{document}